\documentclass[3p,times,twocolumn]{elsarticle}

\usepackage{ecrc}
\usepackage[figuresright]{rotating}
\usepackage{bm,amsmath,amssymb}
\usepackage[mathscr]{eucal}
\usepackage{color}

\biboptions{square,comma,numbers,sort&compress}

\long\def\comment#1{ }
\newcommand{\eqn}[1]{Eq.~\eqref{#1}}
\newcommand{\beq}{\begin{equation}}
\newcommand{\eeq}{\end{equation}}
\newcommand{\nn}{\nonumber\\}
\newcommand{\dif}{{\rm d}}

\newcommand{\rmtr}{{\rm tr}}

\newcommand{\rmJ}{{\rm J}}

\newcommand{\mcal}{\mathcal}

\newcommand{\bk}{\bm{k}}

\newcommand{\bp}{\bm{p}}

\newcommand{\abar}{\bar{\alpha}}

\newcommand{\pd}{{\phantom{\dagger}}}

\newcommand{\Nc}{N_{\rm c}}

\newcommand{\Nf}{N_{\rm f}}

\volume{00}

\firstpage{1}

\journalname{Nuclear Physics B Proceedings Supplement}

\runauth{D.N.~Triantafyllopoulos et al.}

\jid{nppp}

\jnltitlelogo{Nuclear Physics B Proceedings Supplement}

\begin{document}

\begin{frontmatter}




\title{Resumming large higher-order corrections in non-linear QCD evolution}


\author[sac]{E.~Iancu}

\author[sac]{J.D.~Madrigal}

\author[col]{A.H.~Mueller}

\author[sac]{G.~Soyez}

\author[ect]{D.N.~Triantafyllopoulos\corref{cor1}}
\ead{trianta@ectstar.eu}

\address[sac]{Institut de Physique Th\'{e}orique, CEA Saclay, CNRS UMR 3681, F-91191 Gif-sur-Yvette, France}

\address[col]{Department of Physics, Columbia University, New York, NY 10027, USA}

\address[ect]{European Centre for Theoretical Studies in Nuclear Physics and Related Areas (ECT*)\\and Fondazione Bruno Kessler, Strada delle Tabarelle 286, I-38123 Villazzano (TN), Italy}

\cortext[cor1]{Corresponding author}

\begin{abstract}
Linear and non-linear QCD evolutions at high energy suffer from severe issues related to convergence, due to higher order corrections enhanced by large double and single transverse logarithms. We resum double logarithms to all orders by taking into account successive soft gluon emissions strongly ordered in lifetime. We further resum single logarithms generated by the first non-singular part of the splitting functions and by the one-loop running of the coupling. The resulting collinearly improved BK equation admits stable solutions, which are used to successfully fit the HERA data at small-$x$ for physically acceptable initial conditions and reasonable values of the fit parameters. 
\end{abstract}




\end{frontmatter}


\section{Introduction}
\label{sect:intro}

The Color Glass Condensate (CGC) \cite{Gelis:2010nm} is an effective theory for the soft, small-$x$, components of a fast moving hadron. It predicts the saturation of gluon occupation numbers to values of order $\mathcal{O}(1/\abar)$, where $\abar = \alpha_s \Nc/\pi$ with $\Nc$ the number of colors. The most convenient degrees of freedom are Wilson lines for the scattering of projectile partons off the hadron, and gauge invariant combinations of these Wilson lines satisfy unitarity constraints as a result of gluon saturation. These correlators are the building blocks for calculating many observables local in rapidity, like total cross sections in deep inelastic scattering, single and multi-particle production in pA and dA collisions, energy density and its fluctuations in the first stage after an AA collision etc. To an excellent accuracy, such correlators can be expressed in terms of the dipole scattering amplitude $T_{12} = 
	1 - (1/\Nc)\,\rmtr (V_1^{\dagger}V_{2}^{\pd})$ \cite{Dumitru:2011vk,Iancu:2011ns,Iancu:2011nj}. 	
The indices stand for the dependence on the 2-dimensional vectors $z_1$ and $z_2$, which are the transverse coordinates of the quark and the antiquark composing the dipole. The amplitude $T_{12}$ is small in the region where the target is dilute and approaches unity for a dense target. The saturation momentum $Q_s(Y)$ controls the separation between these two regimes, and increases with the rapidity difference $Y$ between the projectile dipole and the hadronic target. 

\section{The BK equation at NLO}
\label{sect:bk}

\begin{figure*}
\begin{center}
\begin{minipage}[b]{0.33\textwidth}
\begin{center}
\includegraphics[width=0.75\textwidth,angle=0]{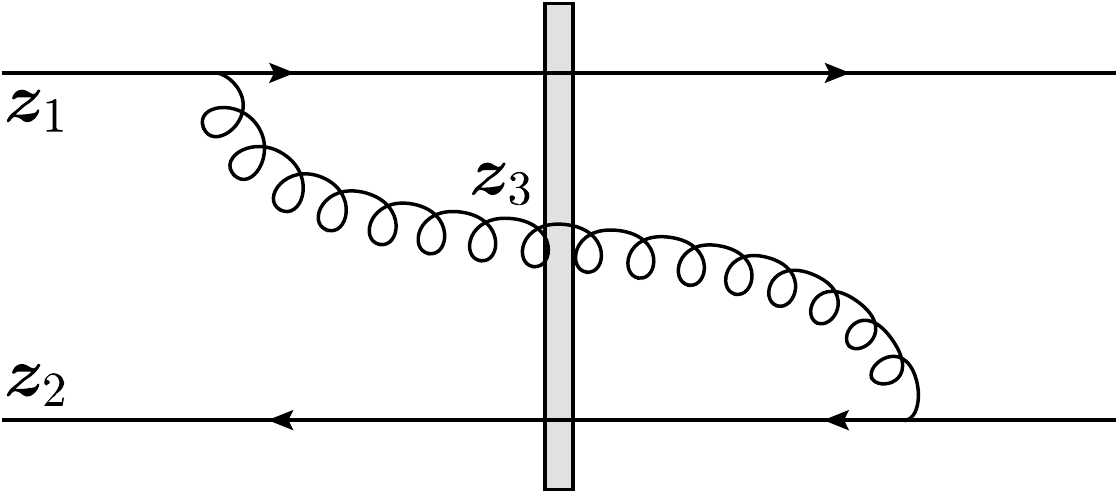}\\(a)\vspace{0.25cm}
\end{center}
\end{minipage}
\begin{minipage}[b]{0.33\textwidth}
\begin{center}
\includegraphics[width=0.75\textwidth,angle=0]{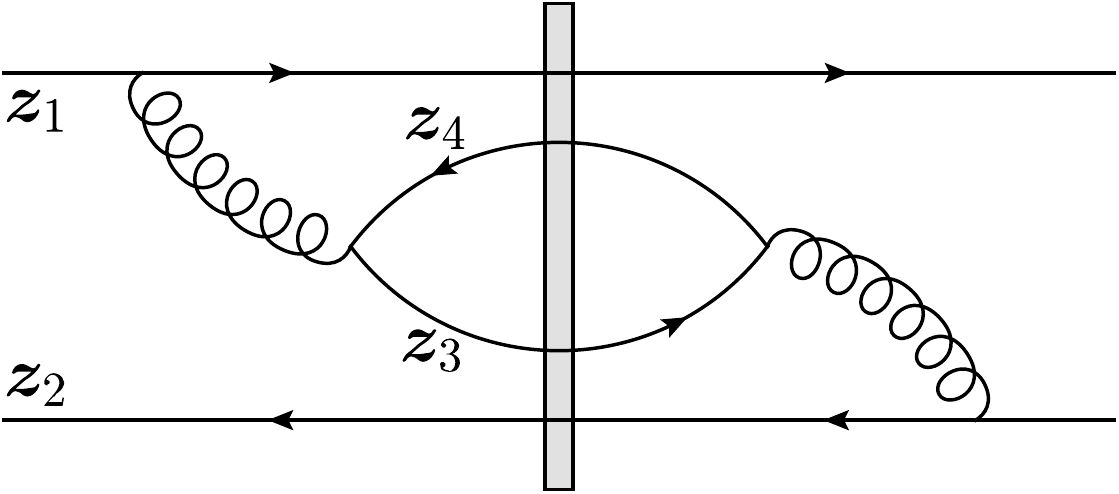}\\(c)\vspace{0.25cm}
\end{center}
\end{minipage}
\begin{minipage}[b]{0.33\textwidth}
\begin{center}
\includegraphics[width=0.75\textwidth,angle=0]{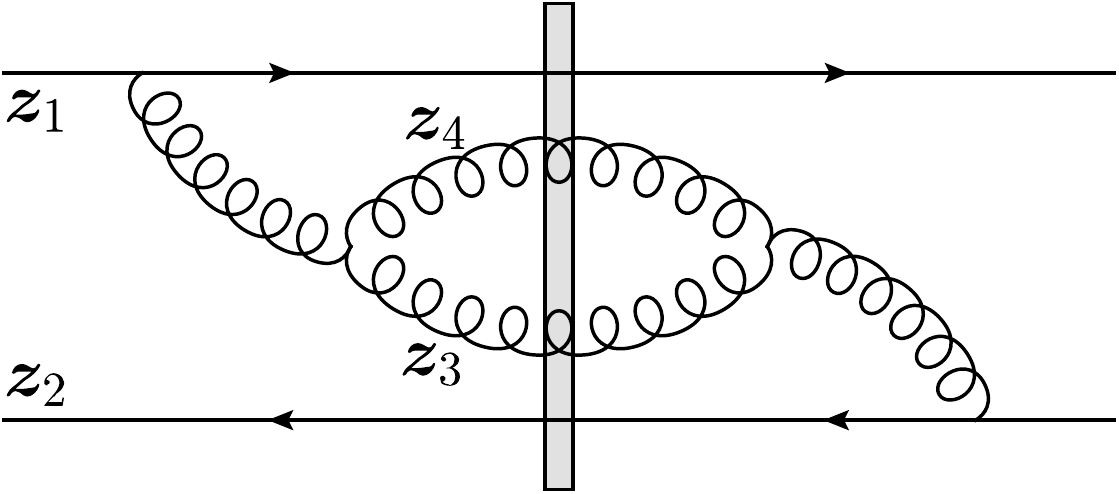}\\(e)\vspace{0.25cm}
\end{center}
\end{minipage}
\begin{minipage}[b]{0.33\textwidth}
\begin{center}
\includegraphics[width=0.75\textwidth,angle=0]{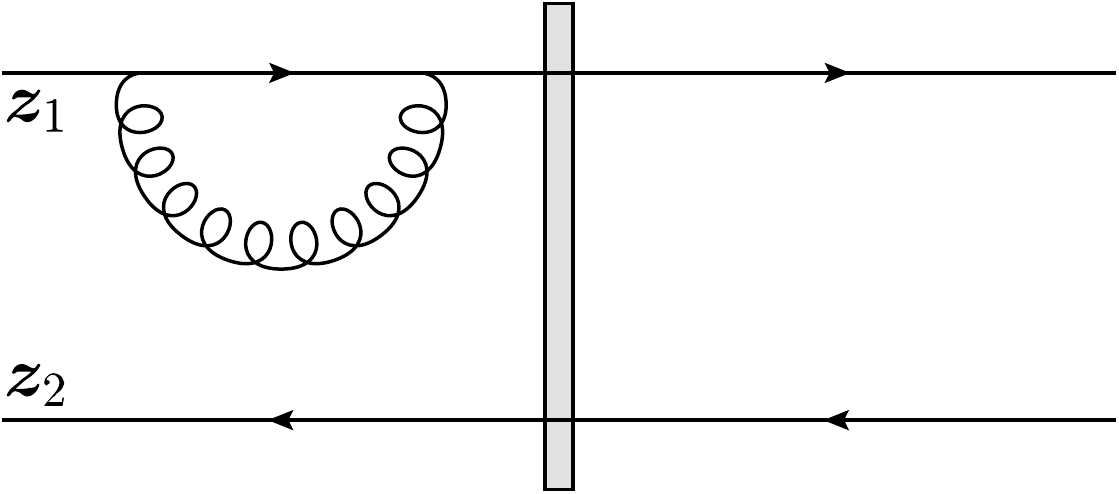}\\(b)\vspace{-0.3cm}
\end{center}
\end{minipage}
\begin{minipage}[b]{0.33\textwidth}
\begin{center}
\includegraphics[width=0.75\textwidth,angle=0]{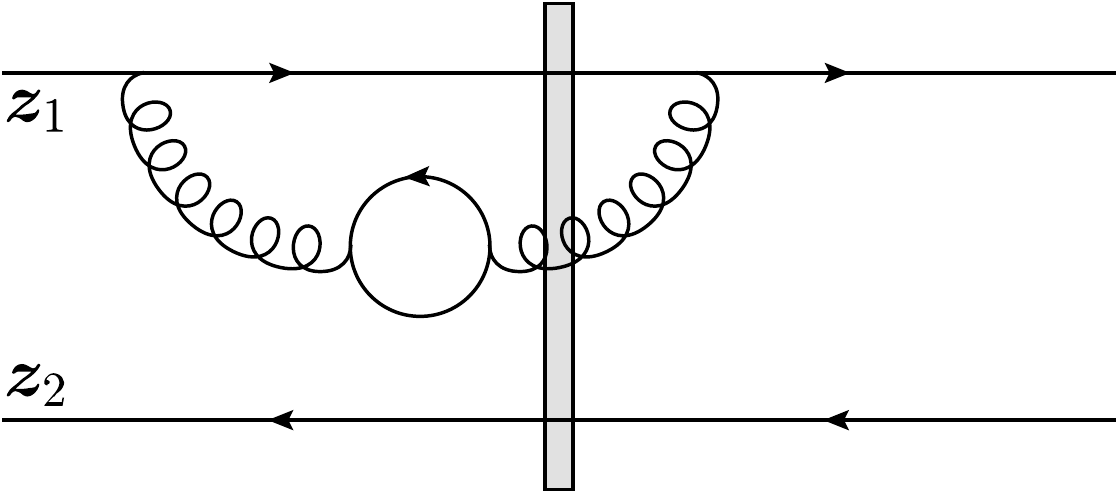}\\(d)\vspace{-0.3cm}
\end{center}
\end{minipage}
\begin{minipage}[b]{0.33\textwidth}
\begin{center}
\includegraphics[width=0.75\textwidth,angle=0]{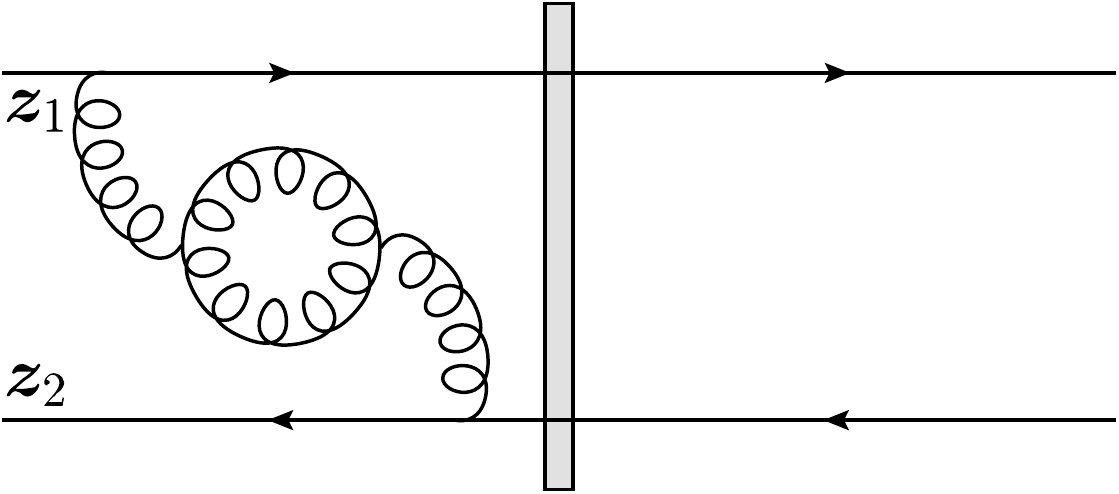}\\(f)\vspace{-0.3cm}
\end{center}
\end{minipage}
\end{center}
\caption{\label{fig:nlodiag} Typical diagrams contributing to the BK equation. The thick vertical line stands for the hadron. Left: LO terms. Middle: NLO terms proportional to $\Nf$. Right: NLO terms proportional to $\Nc$.}
\end{figure*} 

The $S$-matrix $S_{12}=1-T_{12}$ evolves with $Y$ according to the Balitsky-Kovchegov (BK) equation \cite{Balitsky:1995ub,Kovchegov:1999yj}. Typical diagrams contributing to the LO and the NLO evolution involve one or two parton emissions respectively, are shown in Fig.~\ref{fig:nlodiag}, and the corresponding equation written in a form suited to our purposes reads \cite{Balitsky:2006wa,Kovchegov:2006vj,Balitsky:2008zza} 
\begin{align}
 \label{nlobk}
 \hspace*{0cm}
	\frac{\dif S_{12}}{\dif Y}\!= &\frac{\abar}{2\pi}\!
	\int\!\! \dif^2 z_{3}
	\frac{z_{12}^2}{z_{13}^2 z_{23}^2}
	\Bigg[ 1 \!+\! \abar \bar{b}
	\left(\ln z_{12}^2 \mu^2 
 \!-\!\frac{z_{13}^2 \!-\! z_{23}^2}{z_{12}^2}
 \ln \frac{z_{13}^2}{z_{23}^2}\right)
 \nn
 & \hspace*{-1cm} 
 + \abar \left(\frac{67}{36} \!-\! \frac{\pi^2}{12} \!-\! 
 \frac{5}{18}\,\frac{\Nf}{\Nc}
 \!-\! \frac{1}{2}\ln \frac{z_{13}^2}{z_{12}^2} 
 \ln \frac{z_{23}^2}{z_{12}^2}
 \right) \Bigg]
 \left(S_{13} S_{32} \!-\! S_{12} \right)
\nn
 & \hspace*{-1cm}+ \frac{\abar^2}{8\pi^2}
 \int \frac{\dif^2 z_3 \,\dif^2 z_4}{z_{34}^4}
 \Bigg[-2
 + \frac{z_{13}^2 z_{24}^2 + z_{14}^2 z_{23}^2
 - 4 z_{12}^2 z_{34}^2}{z_{13}^2 z_{24}^2 - z_{14}^2 z_{23}^2}
 \nn
 & \hspace*{-0.5cm} 
 \ln \frac{z_{13}^2 z_{24}^2}{z_{14}^2 z_{23}^2}
 +\frac{z_{12}^2 z_{34}^2}{z_{13}^2 z_{24}^2}
 \left(1 + \frac{z_{12}^2 z_{34}^2}{z_{13}^2 z_{24}^2 - z_{14}^2 z_{23}^2} \right)
 \ln \frac{z_{13}^2 z_{24}^2}{z_{14}^2 z_{23}^2}
 \Bigg]
 \nn
 & \hspace*{0.5cm}
 \left(S_{13} S_{34} S_{42}- S_{13} S_{32} \right),
\end{align}
where we have neglected $1/\Nc^2$ suppressed terms\footnote{We have further ignored a term proportional to $\Nf/\Nc$, with $\Nf$ the number of flavors, but which becomes of order $\mathcal{O}(\Nf/\Nc^3)$ in the weak scattering regime.}. In \eqn{nlobk} we have defined $z_{ij}=z_i-z_j$, with $z_3$ and $z_4$ the transverse coordinates of the daughter partons, $\bar{b} = (11\Nc - 2 \Nf)/12\Nc$ is the first coefficient of the QCD $\beta$-function, and $\mu$ is a renormalization scale at which the coupling is evaluated.

\section{Large transverse logarithms}
\label{sect:largelogs}

\begin{figure}
	\begin{center}
	\includegraphics[width=0.33\textwidth,angle=0]{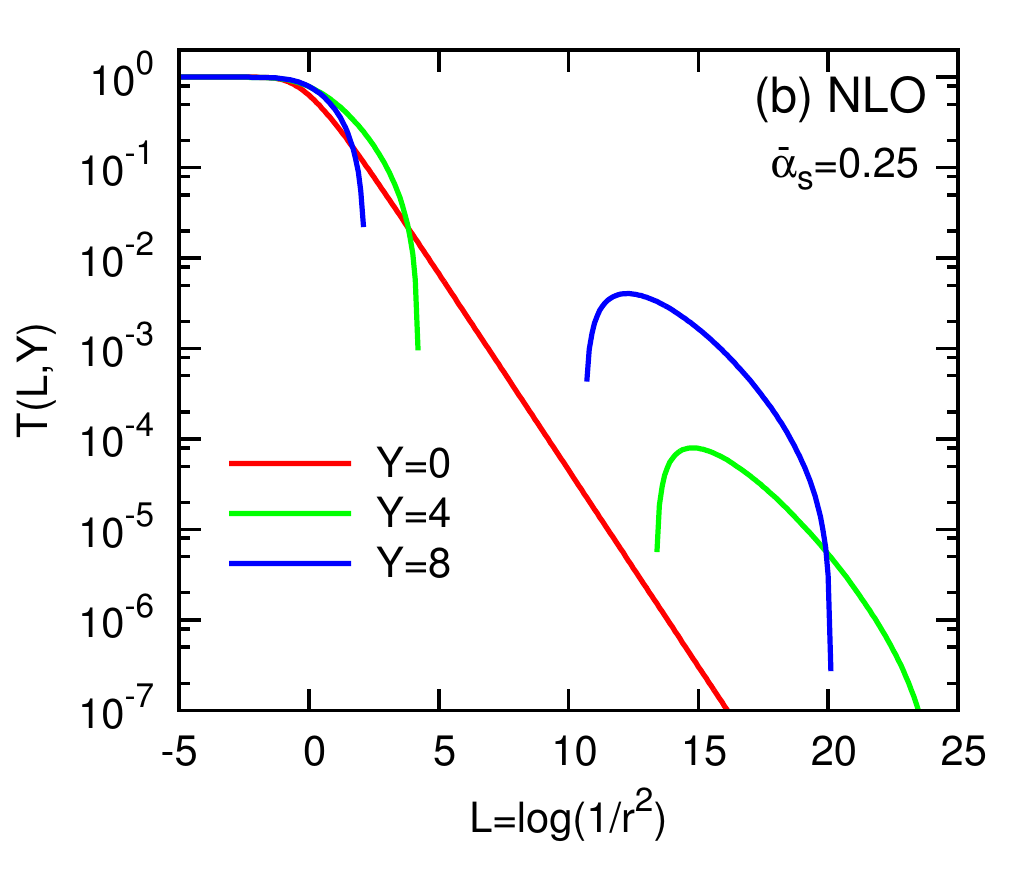}
	\end{center}
    \vspace{-0.5cm}
	\caption{\label{fig:nlodtl} Unstable solution to the NLO BK equation \cite{Iancu:2015vea} . Only the DTL of the NLO kernel is kept and the coupling is fixed at $\abar=0.25$.}
\end{figure}

In general, a NLO computation is expected to add a small $\mathcal{O}(\abar)$ correction to the LO result. However, this is not the case here since there are terms in the NLO kernel which get large in certain kinematic domains and invalidate the strict expansion in $\abar$. These are collinear logarithms, i.e.~logarithms associated with very disparate transverse dipole sizes between successive emissions. More precisely, let's consider the regime
 \begin{equation}
 \label{strord}
 	1/Q_s \gg z_{14} \simeq z_{24} \simeq z_{34}
 	\gg z_{13} \simeq z_{23} \gg z_{12},
 \end{equation}
that is, the parent dipole is the smallest one, a gluon is emitted far away at $z_3$ and another one even further at $z_4$, while all dipoles remain small so that the scattering is weak. Then the dominant NLO contribution involves the double transverse logarithm (DTL) in the single integration term in \eqn{nlobk}. Moreover, a careful expansion of the kernel in the double integration term, reveals that there is also a single transverse logarithm (STL). Linearizing in the amplitude $T$, assuming that the latter depends only on the magnitude of $z_{ij}$, relabelling some variables and putting everything together, we find that \eqn{nlobk} in the collinear region becomes \cite{Iancu:2015vea,Iancu:2015joa}
\begin{equation}
 \label{nlologs}
 	\frac{\dif T_{12}}{\dif Y}
 	= \abar \!\!\!\int\limits_{z_{12}^2}^{1/Q_s^2}\!\!\!
 	\dif z_{13}^2\ \frac{z_{12}^2}{z_{13}^4}
 	\left(1\!-\!
 	\frac{1}{2}\,\abar
 	\ln^2 \!\frac{z_{13}^2}{z_{12}^2} 
 	\!-\! \frac{11}{12}\,\abar
 	\ln \frac{z_{13}^2}{z_{12}^2} \right) T_{13}.
 \end{equation}
To arrive at the above, we have dropped virtual terms, i.e.~terms proportional to $T_{12}$, since largest dipoles scatter stronger. Now it becomes evident that for large enough daughter dipole size $z_{13}$, the NLO contribution can be comparable to, or even larger than, the LO one due to the presence of a large DTL and/or STL. E.g.~consider the simple, but realistic initial condition
\begin{equation}
\label{tin}
	T_{12} = 
	\begin{cases}
 	z_{12}^2 Q_s^2 \quad &\mbox{for} \quad 
 	z_{12}^2 Q_s^2 \ll 1
 	\\
 	1 \quad &\mbox{for} \quad z_{12}^2 Q_s^2 \gg 1.
 \end{cases}
\end{equation}
Then the transverse integration in \eqn{nlologs} becomes logarithmic and the one-step in rapidity evolution gives
 \begin{align}
 \label{dtonestep}
 	\Delta T_{12} = &\, \abar \Delta Y z_{12}^2 Q_s^2\,
 	\ln \frac{1}{z_{12}^2 Q_s^2}
 	\nn
 	&\left( 1 - \frac{1}{6}\,\abar \ln^2\frac{1}{z_{12}^2 Q_s^2}	- \frac{11}{24}\,\abar \ln \frac{1}{z_{12}^2 Q_s^2}
 	  \right).
 \end{align}
Therefore, for sufficiently small $z_{12}$ the perturbation series becomes unreliable and even more, since the large NLO correction is negative, the solution is unstable. This is indeed what has been observed in numerical solutions \cite{Avsar:2011ds,Lappi:2015fma,Iancu:2015vea} and is demonstrated in Fig.~\ref{fig:nlodtl}.

\section{Origin of the large logs and their resummation}
\label{sect:resum}

\begin{figure}
	\begin{center}
	\includegraphics[width=0.30\textwidth,angle=0]{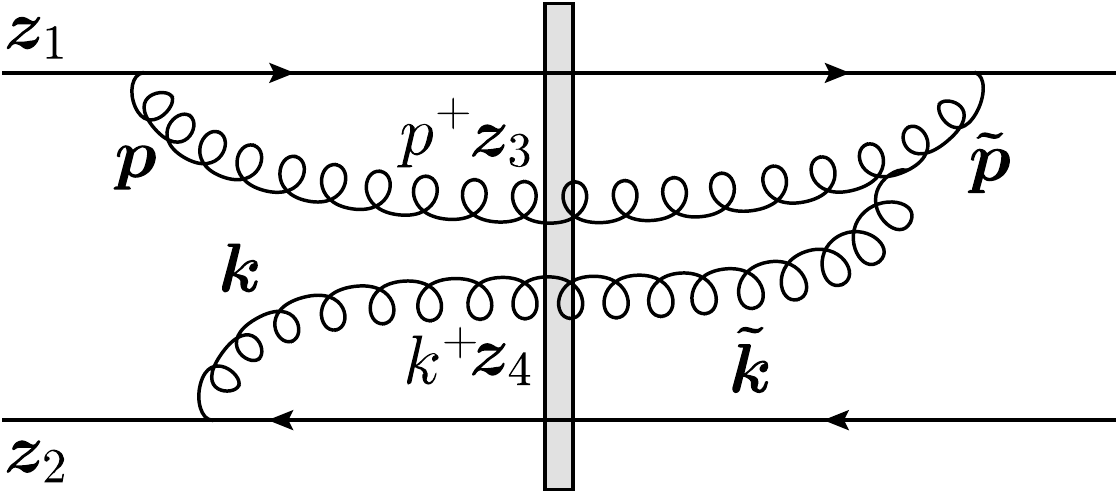}
	\end{center}
	\vspace{-0.3cm}
	\caption{\label{fig:double} Typical diagram leading to a large DTL. The softer gluon $k$ is emitted after and absorbed before the harder one $p$.}
\end{figure}

Understanding the source of the large logarithms is necessary before trying to resum them to all orders. The DTL arises from diagrams in which the two successive gluon emissions are strongly ordered in light-cone time $x^+$, thus its origin is in the kinematics. For example, let us examine the diagram in Fig.~\ref{fig:double} where we take $\bk \ll \bp$, $\tilde{\bk} \ll \tilde{\bp}$ and $k^+ \ll p^+$. Inspection of energy denominators of the type $(k^-+p^-)^{-1} =[(2 \bk^2/k^+) + (2 \bp^2/p^+) ]^{-1}$
$\equiv (1/\tau_k +1/\tau_p)^{-1}$ which appear in light-cone perturbation theory, shows that the largest logarithmic terms occur when the lifetimes (or equivalently the light-cone energies \cite{Beuf:2014uia}) of the two gluons are also strongly ordered, i.e.~when $\tau_k \ll \tau_p$. Different hookings of the two gluons lead to 32 diagrams like the one in Fig.~\ref{fig:double}. Adding all these contributions and integrating over transverse momenta we find in the collinear regime
 \begin{equation}
 \label{deltat12}
 	\Delta T_{12} = \abar^2
 	\!\int\! \frac{\dif p^+}{p^+}\frac{\dif k^+}{k^+}
 	\Theta \left(p^+ \frac{z_{13}^2}{z_{14}^2} -k^+ \right)
 	\dif z_{13}^2 \dif z_{14}^2 \frac{z_{12}^2}{z_{13}^2 z_{14}^4} T_{14}.
 \end{equation}
The step-function is the key element and arises by encoding the lifetime constraint. By further doing the longitudinal integrations and the one over the intermediate dipole size $z_{13}$, we obtain the part with the DTL in \eqn{nlologs}\footnote{Other diagrams which are not time-ordered may contain double logarithms individually, but they cancel in the final answer \cite{Iancu:2015vea}.}. Thus, in order to resum the DTLs to all orders, we need to take into account all the diagrams with an arbitrary number of gluons emissions, in which the gluons are not only strongly ordered in their transverse and longitudinal momenta, but also in their lifetimes.

The STL also arises from successive emissions in which the second gluon is much softer, both in transverse and longitudinal momenta, than the first one, but now it is the region $\tau_k \sim \tau_p$ which gives the relevant contribution. Not surprisingly, the STL is of dynamical origin and is related to DGLAP evolution. This is supported by the fact that its coefficient $A_1=11/12$ is the first non-singular term in the small-$\omega$ expansion of the relevant anomalous dimension, that is 
 \begin{equation}
 	\label{splitting}
 	P_{\rm T}(\omega) = 
 	\int\limits_0^1 \dif z\,z^{\omega}
 	\left[P_{\rm gg}(z)+ 
 	\frac{C_{\rm F}}{\Nc} P_{\rm qg}(z)\right] = 
 	\frac{1}{\omega} - A_1 + \cdots.
 \end{equation}   
This suggests that in order to resum such STLs, it suffices to include $A_1$ as an anomalous dimension, i.e.~as a power-law suppression in the evolution kernel.

Putting everything together, we arrive at the non-linear evolution equation \cite{Iancu:2015vea,Iancu:2015joa}
 \begin{align}
 	\label{bkres}
 	\frac{\dif \tilde{T}_{12}}{\dif Y} = 
 	\frac{\abar}{2\pi}
 	\int \dif^2 z_3 &\,\frac{z_{12}^2}{z_{13}^2 z_{23}^2}
 	\left(\frac{z_{12}^2}{z_{<}^2}\right)^{\pm A_1 \abar}
 	\frac{\rmJ_1\big(2\sqrt{\abar L_{13} L_{23}} \big)}{\sqrt{\abar L_{13} L_{23}}}
 	\nn
 	&\, \big(\tilde{T}_{13} + \tilde{T}_{23} - 
 	\tilde{T}_{12} - \tilde{T}_{13} \tilde{T}_{23}\big),
 \end{align}
which is of the LO BK type, but with two extra factors in the kernel that resum the STLs and DTLs\footnote{So long as the DTLs are considered, a presumably equivalent to the order of accuracy, but non-local in $Y$, equation has been derived in \cite{Beuf:2014uia}. In momentum space and at the linear level, the Bessel kernel has first appeared in \cite{Vera:2005jt}.}. Clearly, when such factors are expanded in the second non-trivial order in $\abar$ they reproduce the large logarithmic terms of the NLO BK equation as presented in \eqn{nlologs}. In \eqn{bkres} we have defined $L_{13} \equiv \ln \big(z_{13}^2/z_{12}^2\big)$ and similarly for $L_{23}$, while $z_{<} =\min\{z_{13},z_{23}\}$. The positive sign in the power-law suppression is to be used in the regime in \eqn{strord}, while the minus sign is taken when $z_{<} \ll z_{12}$ and will resum STLs arising from successive gluon emissions which are still softer in longitudinal momentum but harder in the transverse space. Finally, one should note that $\tilde{T}_{12}$ in \eqn{bkres} coincides with the physical amplitude $T_{12}$ only for $Y \geq \ln(1/z_{12}^2 Q_0^2)$, with $Q_0$ the typical target scale, which is anyway the region in which high energy evolution can be trusted\footnote{Moreover, the initial condition in \eqn{bkres} has to be modified in order to include large transverse logarithms in the impact factor \cite{Iancu:2015vea}.}.

\section{Running coupling}
\label{sect:running}

There is one last source of potentially large NLO terms in the NLO BK equation. These are the logarithmic terms proportional to $\bar{b}$ in \eqn{nlobk} and they can get large when the scales in their arguments become very disparate. The scale $\mu$, which is the same one at which the coupling is to be evaluated in Eqs.~\eqref{nlobk} and \eqref{bkres}, must be chosen in a way to cancel such large logarithms. Even though the choice is not unique, it is the hardest scale which determines the running. Indeed, the smallest dipole prescription
 \begin{equation}
 \label{amin}
 	\bar{\alpha}_{\rm min} = \abar(z_{\rm min})
 	\quad \mbox{with} \quad
 	z_{\rm min} = \min\{z_{12},z_{13},z_{23}\},
 \end{equation}   
cancels the large logarithms in all dangerous regions. Another option is to choose $\mu$ so that all $\abar^2 \bar{b}$ terms vanish. Since in the present work we also neglect all pure $\mcal{O}(\abar^2)$ terms, such a choice looks like the {\em fastest apparent convergence} scheme. One easily finds  
 \begin{equation}
 \label{afac}
 	\bar{\alpha}_{\rm fac} = 
 	\left(\frac{1}{\bar{\alpha}_{12}} + 
 	\frac{z_{13}^2 - z_{23}^2}{z_{12}^2}
 	\frac{\bar{\alpha}_{13} - \bar{\alpha}_{23}}{\bar{\alpha}_{13}\bar{\alpha}_{23}} \right)^{-1},
 \end{equation}
which is equivalent to $\bar{\alpha}_{\rm min}$ in the extreme kinematical cases where one of the three dipoles is much smaller than the other two. We shall adopt one of the above two schemes, but we shall not consider the one introduced in \cite{Balitsky:2006wa} and which has been extensively used so far in phenomenology, as it is plagued by certain peculiarities.

\section{Solutions, applications and outlook}
\label{sect:solution}

\begin{figure}
	\begin{center}
	\includegraphics[width=0.33\textwidth,angle=0]{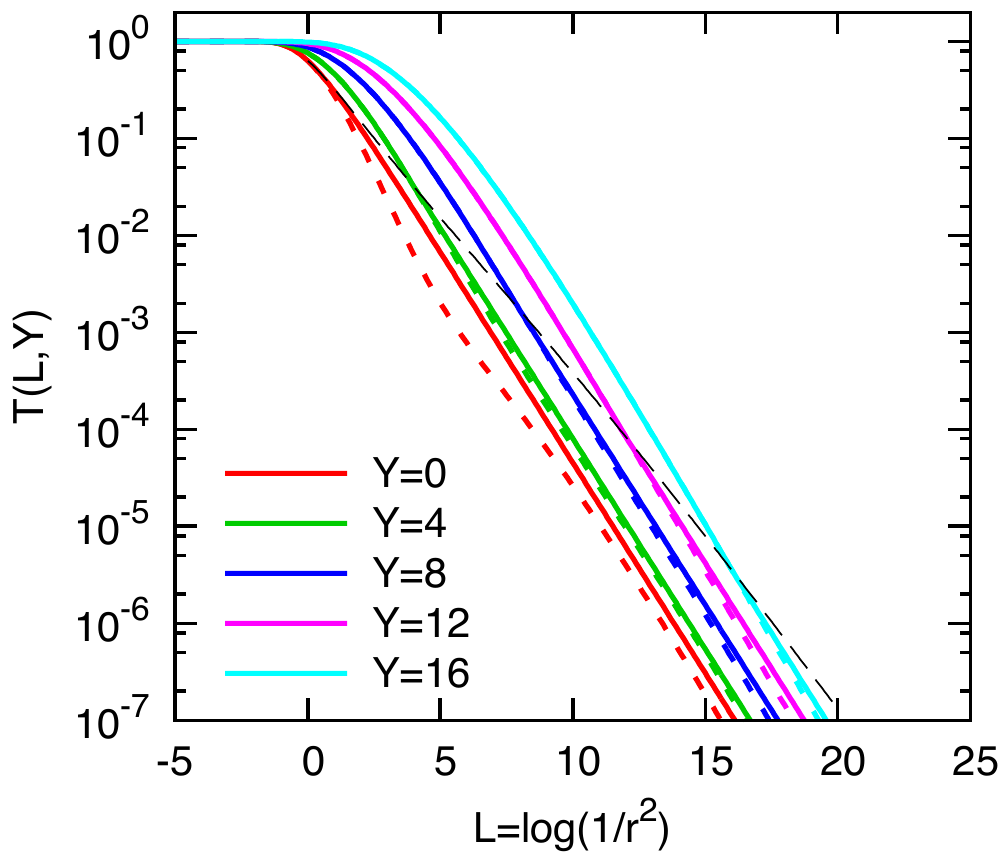}
	\end{center}
	\vspace{-0.5cm}
	\caption{\label{fig:nlores} Stable solution to the resumed BK equation \eqref{bkres}.}
\end{figure}

Contrary to what happens to the NLO BK equation in \eqref{nlobk}, the resummed one in \eqn{bkres} admits well-defined solutions, whether the coupling is fixed or it runs. Such a stable solution is shown in Fig.~\ref{fig:nlores}. The DTLs, the STLs and the running of the coupling individually suppress the evolution and hence the final cumulative ``speed'' is substantially suppressed when compared to the one of the LO evolution.

Given appropriate forms for the initial condition, in \cite{Iancu:2015joa} we have used the numerical solutions to the resummed BK equation in order to perform fits to the HERA data \cite{Aaron:2009aa} for the $ep$ reduced cross section (for similar fits, without inclusion of the STLs, see \cite{Albacete:2015xza}). Given the range of validity for small-$x$ dynamics, we have restricted ourselves in the region $x \leq 10^{-2}$ and $Q^2 \leq Q_{\rm max}^2$, where the upper limit $Q_{\rm max}^2$ in the photon virtuality has been varied even up to 400 GeV$^2$. With 4 free parameters we obtain good fits with a $\chi^2$ per point around 1.1$-$1.2. They are also rather discriminatory: they favor (i) the running coupling version of the McLerran-Venugopalan (MV) model as an initial condition and (ii) the two running coupling prescriptions $\bar{\alpha}_{\rm min}$ and  $\bar{\alpha}_{\rm fac}$, with a slight preference to the former. Whether or not we include the power-law suppression factor due the resummation of STLs in \eqn{bkres}, does not affect the quality of the fits, however it does lead to more physical values of the fit parameters. For completeness, let us point out that the evolution speed extracted from such fits is $\lambda_s \equiv \dif \ln Q_s^2 /\dif Y = 0.20 \div 0.24$. 

It would be desirable to generalize \eqn{bkres} to include the pure $\mcal{O}(\abar^2)$ terms, but this may not change much in the phenomenology, since many of such contributions are taken into account by one of the fit parameters.

\begin{thebibliography}{10}

\bibitem{Gelis:2010nm}
F.~Gelis, E.~Iancu, J.~Jalilian-Marian, R.~Venugopalan, \href{http://dx.doi.org/10.1146/annurev.nucl.010909.083629}{{Ann.Rev.Nucl.Part.Sci.} {60} (2010) 463},
\href{http://arxiv.org/abs/1002.0333}{{arXiv:1002.0333}}.

\bibitem{Dumitru:2011vk}
A.~Dumitru, J.~Jalilian-Marian, T.~Lappi, B.~Schenke, R.~Venugopalan, \href{http://dx.doi.org/10.1016/j.physletb.2011.11.002}{{Phys.Lett.} {B706} (2011) 219},
\href{http://arxiv.org/abs/1108.4764}{{arXiv:1108.4764}}.

\bibitem{Iancu:2011ns}
E.~Iancu, D.N.~Triantafyllopoulos, {JHEP} {1111} (2011) 105,
\href{http://arxiv.org/abs/1109.0302}{{arXiv:1109.0302}}.

\bibitem{Iancu:2011nj}
E.~Iancu, D.N.~Triantafyllopoulos, {JHEP} {1204} (2012) 025,
\href{http://arxiv.org/abs/1112.1104}{{arXiv:1112.1104}}.

\bibitem{Balitsky:1995ub}
I.~Balitsky,
  \href{http://dx.doi.org/10.1016/0550-3213(95)00638-9}{{Nucl.Phys.}
  {B463} (1996) 99},
\href{http://arxiv.org/abs/hep-ph/9509348}{{arXiv:hep-ph/9509348}}.

\bibitem{Kovchegov:1999yj}
Y.V.~Kovchegov,
  \href{http://dx.doi.org/10.1103/PhysRevD.60.034008}{{Phys.Rev.}
  {D60} (1999) 034008},
\href{http://arxiv.org/abs/hep-ph/9901281}{{arXiv:hep-ph/9901281}}.

\bibitem{Balitsky:2006wa}
I.~Balitsky,
  \href{http://dx.doi.org/10.1103/PhysRevD.75.014001}{{Phys.Rev.}
  {D75} (2007) 014001},
\href{http://arxiv.org/abs/hep-ph/0609105}{{arXiv:hep-ph/0609105}}.

\bibitem{Kovchegov:2006vj}
Y.~V. Kovchegov, H.~Weigert, \href{http://dx.doi.org/10.1016/j.nuclphysa.2006.10.075}{{Nucl.Phys.} {A784} (2007) 188},
\href{http://arxiv.org/abs/hep-ph/0609090}{{arXiv:hep-ph/0609090}}.

\bibitem{Balitsky:2008zza}
I.~Balitsky and G.A.~Chirilli, \href{http://dx.doi.org/10.1103/PhysRevD.77.014019}{{Phys.Rev.} {D77} (2008) 014019},
\href{http://arxiv.org/abs/0710.4330}{{arXiv:0710.4330}}.

\bibitem{Iancu:2015vea}
E.~Iancu, J.D.~Madrigal, A.H.~Mueller, G.~Soyez, D.N.~Triantafyllopoulos,
  \href{http://dx.doi.org/10.1016/j.physletb.2015.03.068}{{Phys.Lett.}
  {B744} (2015) 293},
\href{http://arxiv.org/abs/1502.05642}{{arXiv:1502.05642}}.

\bibitem{Iancu:2015joa}
E.~Iancu, J.D.~Madrigal, A.H.~Mueller, G.~Soyez, D.N.~Triantafyllopoulos, \href{http://arxiv.org/abs/1507.03651}{{arXiv:1507.03651}}.

\bibitem{Avsar:2011ds}
E.~Avsar, A.~Stasto, D.N.~Triantafyllopoulos, D.~Zaslavsky,
  \href{http://dx.doi.org/10.1007/JHEP10(2011)138}{{JHEP} {1110}
  (2011) 138},
\href{http://arxiv.org/abs/1107.1252}{{arXiv:1107.1252}}.

\bibitem{Lappi:2015fma}
T.~Lappi, H.~M{\"a}ntysaari,  \href{http://dx.doi.org/10.1103/PhysRevD.91.074016}{{Phys.Rev.}
  {D91} (2015) 074016},
\href{http://arxiv.org/abs/1502.02400}{{arXiv:1502.02400}}.

\bibitem{Beuf:2014uia}
G.~Beuf, \href{http://dx.doi.org/10.1103/PhysRevD.89.074039}{{Phys.Rev.} {D89} (2014) 074039},
\href{http://arxiv.org/abs/1401.0313}{{arXiv:1401.0313}}.

\bibitem{Vera:2005jt}
A.~Sabio~Vera,
  \href{http://dx.doi.org/10.1016/j.nuclphysb.2005.06.003}{{Nucl.Phys.}
  {B722} (2005) 65},
\href{http://arxiv.org/abs/hep-ph/0505128}{{arXiv:hep-ph/0505128}}.

\bibitem{Aaron:2009aa}
{H1, ZEUS} Collaboration, F.~Aaron {et~al.}, \href{http://dx.doi.org/10.1007/JHEP01(2010)109}{{JHEP}
  {1001} (2010) 109},
\href{http://arxiv.org/abs/0911.0884}{{arXiv:0911.0884}}.

\bibitem{Albacete:2015xza}
J.~L. Albacete,
\href{http://arxiv.org/abs/1507.07120}{{arXiv:1507.07120}}.

\end{thebibliography}




\providecommand{\href}[2]{#2}\begingroup\raggedright\endgroup




\end{document}